\RequirePackage{graphicx}
\documentclass[XeLatex,twocolumn,epjc3]{svjour3}          

\usepackage{hyperref}
\usepackage{amsmath}
\usepackage{float}
\usepackage{color}
\usepackage{graphicx}

\begin{document}

\title{Charged strange stellar model with Tolman $V$ metric potential in the Einstein-Maxwell space-time}

\author{Saibal Ray\thanksref{e1,addr1}
\and Dibyendu Shee\thanksref{e2,addr2} \and Debabrata Deb\thanksref{e3,addr3} \and S.K. Maurya\thanksref{e4,addr4} \and M.K. Jasim\thanksref{e5,addr5}.}

\thankstext{e1}{e-mail: saibal@iucaa.ernet.in}
\thankstext{e2}{e-mail: dibyendu\_shee@yahoo.com}
\thankstext{e3}{e-mail: ddeb.rs2016@physics.iiests.ac.in}
\thankstext{e4}{e-mail: sunil@unizwa.edu.om}
\thankstext{e5}{e-mail: mahmoodkhalid@unizwa.edu.om}

\institute{Department of Physics, Government College of Engineering and Ceramic Technology, Kolkata 700010, West Bengal,
India\label{addr1}
\and
Bengal Engineering College Model School, Shibpur, Howrah 711103, West Bengal, India\label{addr2}
\and Department of Physics, Indian Institute of Engineering Science and Technology, Shibpur,
Howrah 711103, West Bengal, India\label{addr3}
\and Department of Mathematics and Physical Science, College of Arts and Science, University of Nizwa, Nizwa, Sultanate of Oman\label{addr4}
\and Department of Mathematics and Physical Science, College of Arts and Science, University of Nizwa, Nizwa, Sultanate of Oman\label{addr5}}

\date{Received: date / Accepted: date}

\maketitle

\begin{abstract}
This paper deals with the existence of a compact stellar object, precisely strange (quark) star,
in the framework of Einstein's General Theory of Relativity with Tolman $V$ metric potential, which is one of the
simplest forms of potential among his proposals. The potential is given by $e^{\nu}=Kr^{2n}$, where $K$ is the
constant and $n$ is a parameter [R.C. Tolman, Phys. Rev. {\bf55}, 364 (1939)]. Considering charged, static,
spherically symmetric, isotropic fluid sphere we have studied different physical features of some strange star candidates namely
$EXO\ 1785-248$, $LMC\ X-4$, $SMC\ X-1$, $SAX\ J1808.4-3658$, $4U\ 1538-52$ and $Her\ X-1$. To represent the strange
quark matter (SQM) distribution we have employed the simplest form of MIT bag equation of state (EOS), which provides a
linear relationship between pressure and density of the matter through Bag constant $B$. We have done several tests for the stability criteria
and physical acceptability of the proposed model. The results show consistency with energy condition, TOV equation, adiabatic index, etc. 
We have calculated different physical parameters of our model for the three different consecutive values of
 Bag constant $B$ which are $83\ MeV/fm^3$, $90\ MeV/fm^3$ and $100\ MeV/fm^3$.
Among them with $B=90\ MeV/fm^3$ we have analyzed different properties of the proposed strange star candidates.
\end{abstract}


\maketitle

\section{Introduction}
Albert Einstein, a creator of General Theory of Relativity~\cite{Einstein1915}, is still glowing with his own charm like a Sun in the sky. His creativity changed the definition of Theoretical Astrophysics and Cosmology~\cite{Connor1996}. Overcoming the pillars of Newtonian gravity, his creation makes a revolution in the space-time of physics.  Though scientists observed his proficiency regarding relativity during Special Theory of Relativity in $1905$~\cite{Einstein1905}. Actually, it opened up the hidden mysteries of the universe to us. 

We know that type II supernovae explosion of massive star $(M>8M_\odot)$ gives birth of neutron stars. Though this concept gets concrete support after the observational detection of pulsars~\cite{Hewish1968}. Extremely dense neutron stars can distort the space-time geometry. Their masses lies between $(1.4-2)~M_\odot$ and radius remains between (11-15) km~\cite{Demorest2010}. According to Duncan and Thompson~\cite{Duncan1992} a strong magnetic field exists in such compact objects. Some review also shows that a strong magnetic field of the order of $10^{14}-10^{15}$ Gauss exists near the surface of highly compact magnetars, strongly magnetized neutron stars~\cite{Thompson1996,Ibrahim2002}, which is its chief characteristic features. But according to Chakraborty~\cite{Chakraborty1997} core of the neutron star also contains an even much stronger magnetic field of the order of $10^{19}-10^{20}$ Gauss. The reason behind the development and existence of this strong magnetic field is still unknown to all of us. But quark spins~\cite{Tatsumi2000} and spontaneous ordering of nucleon~\cite{Isayev2004,Isayev2006} may be taken as a reason for this magnetic field. The core of the neutron star possesses extreme pressure and high density, as a result, the interior of the neutron star suffers from a phase transition of neutrons to hyperon and strange quark matter (SQM). According to Cameron~\cite{Cameron1959} production of hyperon from
neutron is energetically more favorable due to weak interaction among them and extremely high density.  Quark matter is formed from the quarks of nucleons to form a colorless matter. The quark star contain equal number of up ($u$), down ($d$) and strange ($s$)
quarks~\cite{Bodmer1971,Witten1984,Terazawa1989,Terazawa1990,Alcock1986,Itoh1970,Haensel1986,Farhi1984}. Due to this extreme density hyperons are deconfined into strange quarks. Among the up ($u$), down ($d$) and strange ($s$) quarks, the most stable one is strange quarks. Once the up ($u$) and down ($d$) quarks of the quark star converted into strange matter the entire quark matter gets converted into strange matter, since the strange matter is the true ground state of nuclear matter~\cite{Pagliara2013,Alcock1988,Madsen1999}.
According to Chodos et al.~\cite{Chodos1974} a finite region of space confined with fields can be considered as a strongly interacting particle. So he considers the Bag constant $(B)$ which is a finite region with constant energy density. This constant must affect the energy-momentum tensor (i.e. geometry of the space-time) of the star. According to the MIT bag model, the reason for quark confinement is due to the universal Bag constant $(B)$. This constant actually represents the difference between the perturbative and nonperturbative Quantum Chromodynamics Vacuum (QCD), in terms of energy density. According to Farhi~\cite{Farhi1984} and Alcock~\cite{Alcock1986} (55-75) MeV/fm$^3$ should be range of the value of the Bag constant. Though CERN-SPS and RHIC proposed wide range values of the Bag constant~\cite{Burgio2002}.

The study of the charged fluid sphere, static in nature, in general relativity, is an issue of a challenge because its presence creates an electric repulsion in addition to the thermal pressure gradient of fluid which averts the gravitational collapse~\cite{Felice1995,Sharma2001}. According to Ivanov~\cite{Ivanov2002}, consideration of charged perfect fluid sphere prevent the growth of curvature of space-time, as a result, we can avoid singularities. Bonnor~\cite{Bonnor1965} considers a dust distribution with a small amount of charge. Due to the repulsive force of this charge, the dust distribution remains in equilibrium against the gravitational pull. Das et al.~\cite{Das2011} studied isotropic static charged fluid spheres with two different specializations in Einstein-Maxwell space-time. Malaver~\cite{Malaver2014} studied static, spherically symmetric, charged quark stars through Tolman $IV$ metric potential and quadratic equation of state. According to his consideration charge density becomes singular otherwise physical analysis is well behaved. Ray et al.~\cite{Ray2003} studied the effect of electric charge on the compact star which is proportional to its mass density. They studied Einstein-Maxwell field equations through polytropic EOS and concluded that the charged star can collapse to form a charged black hole before all the charge leaves the system. Negreiros et al.~\cite{Negreiros2009} studied electrically charged strange quark star with the possible existence of ultra-strong electric fields on their surface which is of the order of $10^{18}\ V/cm$. The electric field may exceed the value of $10^{19}\ V/cm$ if the strange matter forms a color superconductor.

 Arba\~{n}il  and Malheiro~\cite{Arbanil2015} studied the equilibrium and stability of a charged strange quark star considering MIT bag model equation of state and the radial charge distribution follows a power law.  Paulucci and Horvath~\cite{Paulucci2014} presented beautifully that, in high-temperature astrophysical events, the fragmentation of Strange Quark Matter takes place into strangelets. They have done it with the consideration of MIT bag model framework in color-flavor-locked (CFL) state~\cite{Alford1999,Rapp2000,Lugones2002}.

 An ultra-dense strange (quark) stars with the MIT bag model and Mak and Harko~\cite{Mak2002} density profile was studied by Deb et al.~\cite{Deb2017}. To this end, they showed the typical mass-radius relation and featured a very unique result that anisotropy of compact strange star increases with the radius of the star and attains a maximum value at the surface. Deb et al.~\cite{Deb2018a} also studied the generalization of the above case with charge distribution of the given form $q(r)=\alpha r^3$, where $\alpha$ is a constant. In this article, they provided anisotropic charged, a spherically symmetric stellar model that is suitable to study ultra-dense strange stars. Jasim et al.~\cite{Jasim2018} studied anisotropic strange stars, with massive strange quarks whose EOS is $p_r=0.28(\rho-4B)$ and Cosmological constant $\Lambda$, in Tolman-Kuchowicz (TK) space-time. They predicted exact values of the radius of the different strange stars and the solutions satisfy all the physical requirements of stability. Interestingly more the value of $\Lambda$ and $B$ make the system more compact ultra-dense stellar object. Also there are so many literature survey available regarding the MIT bag model EOS and strange
stars~\cite{Glendenning1995a,Glendenning1995b,Kettner1995,Alford2008,Weber2005,Glendenning2000,Weber1999,Sedrakian2007,Golf2009,Glendenning1990,Brilenkov2013,Panda2015,Bhar2015}. In this connection, some other exact models have been investigated without EOS in general relativity as well as modified $f(R,T)$ gravity theory \cite{m1,m11,m12,m13,m2,m3,m4,m5,m6,m7}. 
Motivated by one of the earlier work of Shee et al.~\cite{Shee2018} we present here the charged generalization of it. In that paper, they present a strange star model, with Tolman $V$~\cite{Tolman1939} metric potential and MIT bag equation of state, made off with an anisotropic fluid. They studied the model for different values of $n$, where $n$ is the parameter of Tolman $V$ metric potential, but $n=\frac{1}{2}$ gives a physically acceptable solution as in the original paper Tolman assumed. Through various physical tests, the model presents a stable configuration.
Using the above concept, we have studied charged and isotropic fluid sphere under Tolman $V$ metric potential. The exact solutions of Einstein-Maxwell field equations lead us to understand the physical nature of the strange stars, by incorporating typical mass-radius relationship of them. The singular nature of the pressure and density makes the situation more crucial and interesting to study our case deeper than any other ultra-dense compact stellar system. But tests of the physical validity of our model provide positive feedback to go ahead. We have studied our model for different strange star candidates namely $EXO\ 1785-248$, $LMC\ X-4$, $SMC\ X-1$, $SAX\ J1808.4-3658$, $4U\ 1538-52$ and $HER\ X-1$ and for different values of the bag constant $(B)$ like $83\ MeV/fm^3$, $90\ MeV/fm^3$ and $100\ MeV/fm^3$. Inspired by the literature of Rahaman et al.~\cite{Rahaman2014} and Kalam et al.~\cite{Kalam2013} we have chosen the above values of the bag constants.

The above literature survey motivates us to perform the present research work on the following line of action. In Sect. 2 the basic mathematical structure of Einstein-Maxwell spacetime with isotropy and electric field has been provided by us. Sect. 3 contains solutions to the field equations and other parameters of the astrophysical scenario. To represent the model properly we have found out the constants with respect to the known parameters in Sect. 4. The physical features of our model are represented in Sect. 5 through various subsections namely, Energy Conditions, TOV equation, Herrera's Causality condition, Adiabatic Index, Mass radius relationship and redshift of the isotropic charged strange star. Lastly, in Sect. 6 some concluding remarks and discussions are made by us to open the hidden mysteries and different aspect of our strange star model.

\section{Mathematical Structure of Einstein-Maxwell Space Time}
Let us consider the metric~\cite{Hobson2006} (also note the metric and allied assumptions made in the following works~\cite{Thirukkanesh2008,Negreiros2009,Malheiro2011,Rahaman2012,Sunzu2014,Panahi2016,Arbanil2015}) to describe the curvature coordinate of ultra dense spherically symmetric stellar system
\begin{equation}
{ds}^{2}={{e}^{\nu(r)}}{{dt}^{2}}-{{e}^{\lambda(r)}}{{dr}^{2}}-{r}^{2}({{d\theta}^{2}}+{{sin}^{2}}\theta{{d\phi}^{2}}),\label{eq1}
\end{equation}
where the metric potentials $\nu(r)$ and $\lambda(r)$ are the function of radial coordinate only. These metric coefficients have much significance to understand the gravitational environment of the stellar model. The signature of the space time is taken here $(+,-,-,-)$. Now the Einstein-Maxwell field equations for obtaining the hydrostatic stellar structure of the charged sphere can be provided as
\begin{equation}
G^{i}_{j}=\mathcal{R}^{i}_{j}-\frac{1}{2}\mathcal{R}\delta^{i}_{j}=-\kappa(T^{i}_{j}+E^{i}_{j}),\label{eq2}
\end{equation}
where we chosen $G=1=c$ in the relativistic geometrized unit so that $\kappa(=8\pi)$ is the Einstein's constant. In the above equation $\mathcal{R}^{i}_{j}$ and $\mathcal{R}$ represents Ricci Tensor and Ricci Scalar respectively. $T^{i}_{j}$ and $E^{i}_{j}$ are the energy momentum tensor of the perfect fluid and electromagnetic field are defined as
\begin{equation}
T^{i}_{j}=(\rho+p)v^{i}v_{j}-p\delta^{i}_{j},\label{eq3}
\end{equation}
\begin{equation}
E^{i}_{j}=\frac{1}{4\pi}(-F^{im}F_{jm}+\frac{1}{4}\delta^{i}_{j}F^{mn}F_{mn}),\label{eq4}
\end{equation}
where $p$, $\rho$, $v^{i}$ and $F_{ij}$ denotes fluid pressure, energy density, four velocity vector and anti symmetric electromagnetic field strength
tensor which can be defined as
\begin{equation}
F_{ij}=\frac{\partial A_{j}}{\partial x_{i}}-\frac{\partial A_{i}}{\partial x_{j}},\label{eq5}
\end{equation}
where, $ A_{j}=(\phi(r),0,0,0)$ is the four potential. $F_{ij}$ satisfies the covariant equations of Maxwell,
\begin{equation}
F_{ik,j}+F_{kj,i}+F_{ji,k}=0,\label{eq6}
\end{equation}
\begin{equation}
[\sqrt{-g} F^{ik}],_{k}=-4\pi J^{i}\sqrt{-g},\label{eq7}
\end{equation}
where the electromagnetic four current vector $J^{i}$ is defined as
\begin{equation}
 J^{i}=\frac{\sigma}{\sqrt{g_{44}}} \frac{dx^i}{dx^4}=\sigma v^i,\label{eq8}
\end{equation}
where $\sigma=e^{\frac{\nu}{2}}J^{0}$ represents the charge density and $g$ is the determinant of the metric $g_{ij}$ defined by
\begin{equation}
g=-e^{\nu+\lambda}r^{4}{sin}^{2}\theta,\label{eq9}
\end{equation}
and $J^{0}$ is the only non vanishing component of the electromagnetic four current $J^{i}$ for a static spherically symmetric stellar system.

The total charge, following the relativistic Gauss's law, with a sphere of radius $r$ can be defined as
\begin{equation}
q(r)=r^{2}E(r)=4\pi\int_{0}^{r}J^{0}r^{2}e^\frac{\nu+\lambda}{2}dr,\label{eq10}
\end{equation}
where $E(r)$ being the intensity of the electric field. Using the above mathematical background  we therefore have the Einstein-Maxwell field equations for the isotropic charged spherically symmetric stellar system as follows~\cite{Dionysiou1982,Nduka1976},

\begin{eqnarray}
&\qquad\hspace{-2.4cm} {{\rm e}^{-\lambda }}\left( {\frac {\lambda^{{\prime}}}{r}}-\frac{1}{r^2}\right)+\frac{1}{r^2}=8\pi  \rho+{E}^{2},\label{eq11}\\
&\qquad\hspace{-2.4cm} {{\rm e}^{-\lambda}} \left( {\frac {\nu^{{\prime}}}{r}}+\frac{1}{r^2}\right) -\frac{1}{r^2}=8\pi  p-{E}^{2},\label{eq12}\\
&\qquad\hspace{-1.0cm} \frac{{\rm e}^{-\lambda}}{2} \left( \nu^{{\prime\prime}}+\frac{{\nu^{{\prime}}}^{2}}{2}+{\frac {\nu^{{\prime}}-\lambda^{{\prime}}}{r}}-\frac{\nu^{{\prime}}\lambda^{{\prime}}}{2} \right) =8\pi p+{E}^{2},\label{eq13}
\end{eqnarray}
along with the conservation equation
\begin{equation}
-\frac{(\rho+p)\nu'}{2}-\frac{dp}{dr}+\frac{q}{4\pi\,r^4}\,\frac{dq}{dr}=0. \label{eq14a}
\end{equation}
where`$\prime$' and `$\prime\prime$' denotes the derivative and double derivative with respect to the radial parameter $r$.

Since Einstein-Maxwell field equations are highly nonlinear in nature, they are not easily solvable. Some well behaved model of charged perfect fluid relativistic matter, with proper analytical solutions are given in the following literatures~\cite{Stephani2003,Sabbadini1973,Glass1978,Hartle1978,Durgapal1984,Durgapal1985,Sarracino1985,Delgaty1998}. In this paper we want to study Einstein-Maxwell field equations using Tolman $V$ metric potential, which is given by
\begin{equation}
e^{\nu}=Kr^{2n},\label{eq14}
\end{equation}
where $K$ is a constant and $n$ being a parameter. Tolman originally chosen its value $\frac{1}{2}$ for physically valid solutions.

To describe SQM distribution of strange star we consider the most simple MIT bag model EOS. According to this model, the quarks are massless and noninteacting, the quark pressure can be represented by the following relation, where all the corrections due to pressure and energy are included, can be defined as~\cite{Alcock1986,Farhi1984,Chodos1974},
\begin{equation}
p=\sum_{f=u,d,s} p^{f}-B,\label{eq15}
\end{equation}
where $p^{f}$ represents the pressure due to individual quark flavors. $B$ is the `Bag constant' represents vacuum energy density. The
quark pressure and energy density of each quark flavor are related by the following relation $p^{f}=\frac{1}{3}\rho^{f}$. So, the energy density of the
deconfined SQM distribution according to this model is given by
\begin{equation}
\rho=\sum_{f=u,d,s} \rho^{f}+B,\label{eq16}
\end{equation}
using Eq. (\ref{eq15}) and Eq. (\ref{eq16}) we have MIT bag EOS as
\begin{equation}
p=\frac{1}{3}(\rho-4B).\label{eq17}
\end{equation}
This simple model is very useful to study equilibrium configuration of a ultradense compact object in the general relativity as well as in modified
gravity \cite{mauryaprd,mauryaepjc1}, without involving quantum mechanical particle aspect.

The exterior space time of our charged system can be described by the Reissner-Nordstr{\"o}m metric~\cite{Reissner1916,Nordstrom1918} which is given as
\begin{eqnarray}
&\qquad ds^2 = \left(1 - \frac{2M}{r} +\frac{Q^2}{r^2}\right) dt^2- \frac{1}{\left(1 - \frac{2M}{r} + \frac{Q^2}{r^2}\right)} dr^2 \nonumber  \\
&\qquad - r^2(d\theta^2 + sin^2\theta d\phi^2),\label{eq18}
\end{eqnarray}
where $Q$ is the total charge within the boundary of the star.

\section{Solutions to the Einstein-Maxwell Field Equations}
Using Eq. (\ref{eq11})- Eq. (\ref{eq14}), Eq. (\ref{eq17}) and Eq. (\ref{eq18}) we get the solution of the Einstein field equations given as follows:
\begin{eqnarray}
&\qquad\hspace{-0.8cm} \rho=[\rho_{{1}} \left( n+\frac{1}{2}\right) {r}^{{\frac {-2\,{n}^{2}-2\,n-2}{n+2}}}+16\, \left( n+2 \right)  \nonumber \\
&\qquad\hspace{-1.0cm}\left ( B\pi \,{r}^{2}{n}^{4}+\frac{3}{16}{n}^{3}+ \left( 6\,B\pi \,{r}^{2}+\frac{3}{8} \right) {n}^{2}\right.\nonumber \\
&\qquad\hspace{-1.8cm} \left.+ \left( 5\,B\pi \,{r}^{2}+{\frac {9}{16}} \right) n+6\,B\pi \,{r}^{2}) \right]\Bigg/  \nonumber \\
&\qquad\hspace{-0.6cm}[16\, \left( {n}^{2}+2\,n+3 \right)  \left( n+2 \right)  \left( {n}^{2}+n+1 \right) \pi \,{r}^{2}],\label{eq19}
\end{eqnarray}

\begin{eqnarray}
&\qquad\hspace{-1.4cm} p=[\rho_{{1}} \left( n+\frac{1}{2}\right) {r}^{{\frac {-2\,{n}^{2}-2\,n-2}{n+2}}}-48\, \left( n+2 \right) \nonumber \\
&\qquad\hspace{-0.5cm}\left( B\pi \,{r}^{2}{n}^{4}+ \left( 4\,B\pi \,{r}^{2}-\frac{1}{16} \right) {n}^{3}+ \left( 6\,B\pi \,{r}^{2}-\frac{1}{8}\right) {n}^{2}\right.\nonumber \\
&\qquad\hspace{-1.8cm} \left.+\left( 5\,B\pi \,{r}^{2}-\frac{3}{16} \right) n+2\,B\pi \,{r}^{2} \right)] \Bigg/ \nonumber \\
&\qquad\hspace{-0.6cm}[48\, \left( {n}^{2}+2\,n+3 \right)  \left( n+2 \right)  \left( {n}^{2}+n+1 \right) \pi \,{r}^{2}],\label{eq20}
\end{eqnarray}

\begin{eqnarray}
&\qquad\hspace{-1.4cm} {E}^{2}=[-\rho_{{1}} \left( n+3 \right) {r}^{{\frac {-2\,{n}^{2}-2\,n-2}{n+2}}}-48\,n \left( n+1 \right) \nonumber \\
&\qquad\hspace{-0.5cm}  \left( n+2 \right)  (  \left( B\pi \,{r}^{2}-\frac{1}{8}\right) {n}^{3}-\frac{3}{16}\,{n}^{2}-\frac{n}{4}-B\pi \,{r}^{2}+\frac{3}{16} )] \Bigg/ \nonumber\\
&\qquad\hspace{-1.0cm}[6\, \left( n+1 \right) \left( n+2 \right)  \left( {n}^{2}+2\,n+3 \right) \left( {n}^{2}+n+1 \right) {r}^{2}],\label{eq21}
\end{eqnarray}

\begin{eqnarray}
&\qquad\hspace{-1.0cm} \lambda=ln\Bigg[\Big\lbrace(n^2+n+1)\left( {n}^{2}+2\,n+3 \right) {r}^{{\frac {2\,{n}^{2}+2\,n+2}{n+2}}} \Big\rbrace \Bigg/ \nonumber \\
&\qquad\hspace{-0.5cm}\Big\lbrace \Big(  \left( -16\,B\pi \,{r}^{2}+1 \right) {n}^{2}+ \left( -16\,B\pi \,{r}^{2}+2 \right) n \nonumber \\
&\qquad\hspace{-0.1cm} -16\,B\pi \,{r}^{2}+3 \Big) {r}^{{\frac {2\,{n}^{2}+2\,n+2}{n+2}}} + \Big( 16\,B\pi \, \left( {n}^{2}+n+1 \right) {R}^{4} \nonumber \\
&\qquad\hspace{-1.8cm} +n \left( n+1 \right)  \left( {n}^{2}+2\,n+3 \right) {R}^{2}\nonumber \\
&\qquad\hspace{-1.2cm} -2\,M \left( {n}^{2}+n+1 \right)  \left( {n}^{2}+2\,n+3 \right)R\nonumber \\
&\qquad\hspace{-0.8cm}+{Q}^{2} \left( {n}^{2}+n+1 \right)  \left( {n}^{2}+2\,n+3 \right) \Big) {R}^{{\frac {2\,{n}^{2}-2}{n+2}}} \Big\rbrace\Bigg],\label{eq22}
\end{eqnarray}
where
\begin{eqnarray}
&\qquad\hspace{-1.0cm} \rho_{{1}}=96\, \left( n+\frac{1}{2} \right) \left( n+1 \right)  \Bigg[ \frac{1}{16}\,n \left( n+1 \right)\nonumber \\
&\qquad\hspace{-0.4cm} \left( {n}^{2}+2\,n+3 \right) {R}^{{\frac {2\,{n}^{2}+2\,n+2}{n+2}}}+
 \Big\lbrace {R}^{{\frac {2\,{n}^{2}+4\,n+6}{n+2}}}B\pi \nonumber \\
&\qquad\hspace{-1.8cm}  -\frac{1}{8}\, \left( -\frac{1}{2}\,{R}^{{\frac {2\,{n}^{2}-2}{n+2}}}{Q}^{2}+{R}^{{\frac {2\,{n}^{2}+n}{n+2}}}M \right) \nonumber \\
&\qquad\hspace{-1.8cm}\left( {n}^{2}+2\,n+3 \right)  \Big\rbrace  \left( {n}^{2}+n+1 \right)  \Bigg].\label{eq23}
\end{eqnarray}


\begin{figure}[!htpb]\centering
\includegraphics[width=6cm]{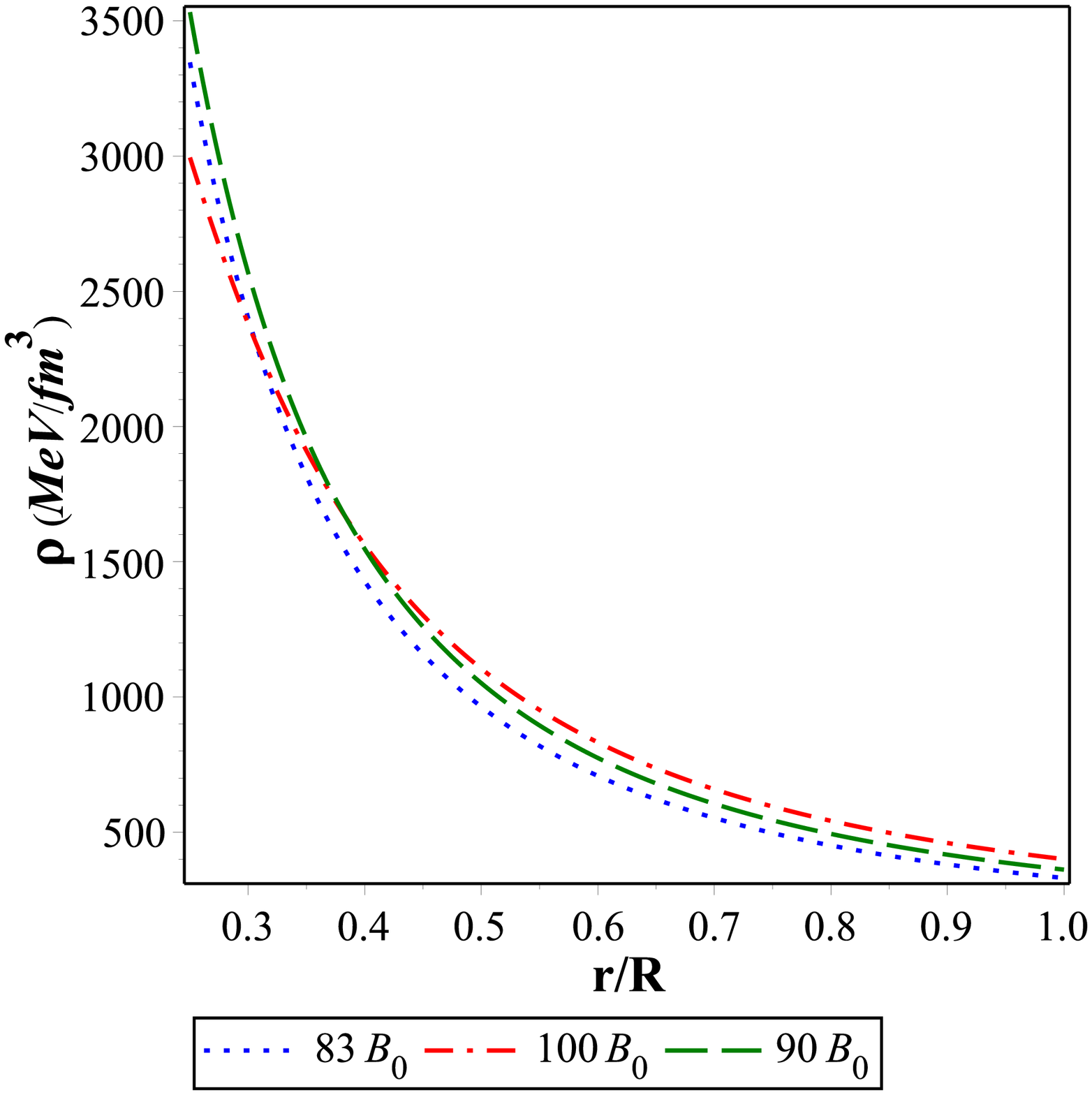}
\includegraphics[width=6cm]{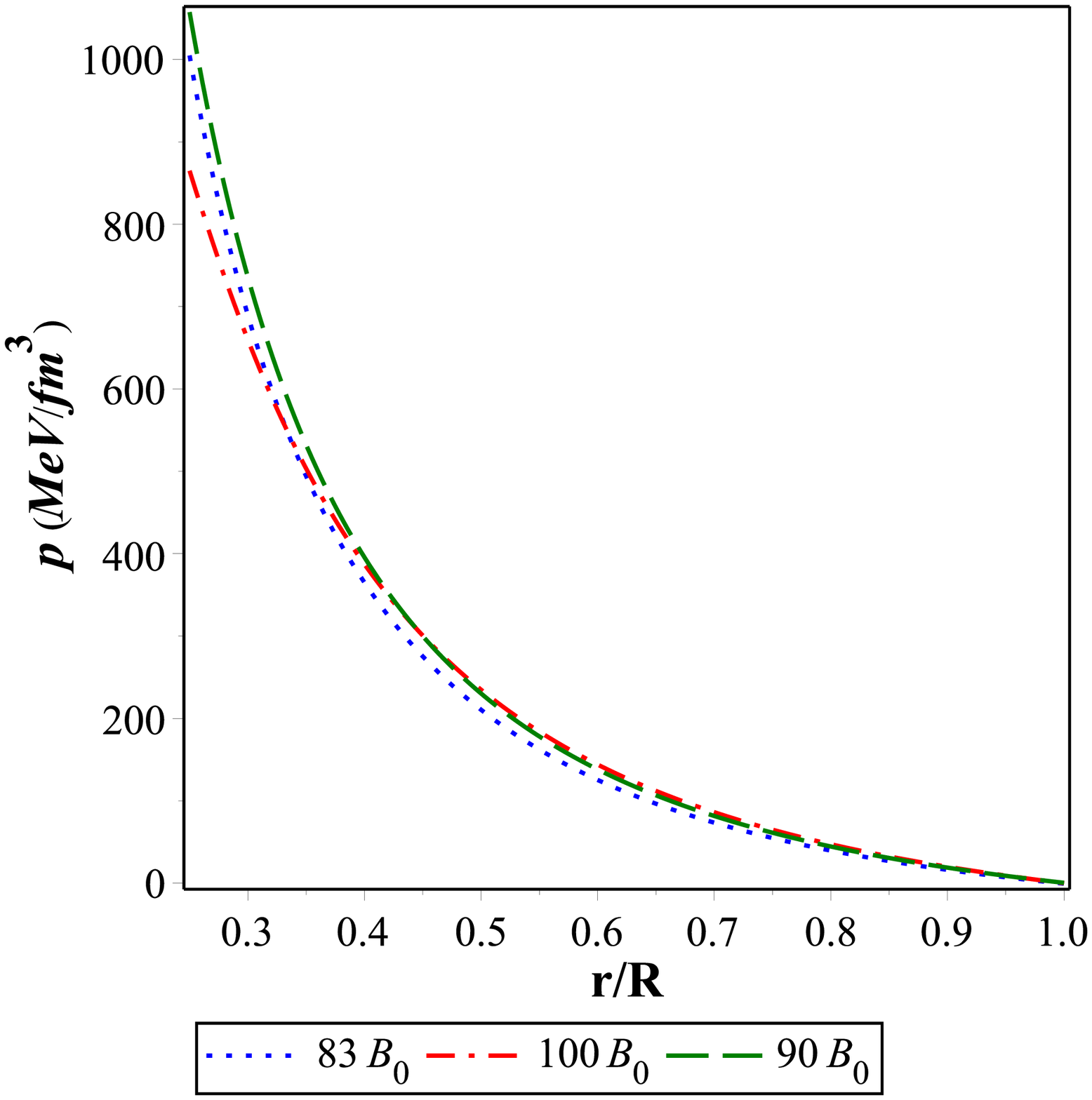}
\includegraphics[width=6cm]{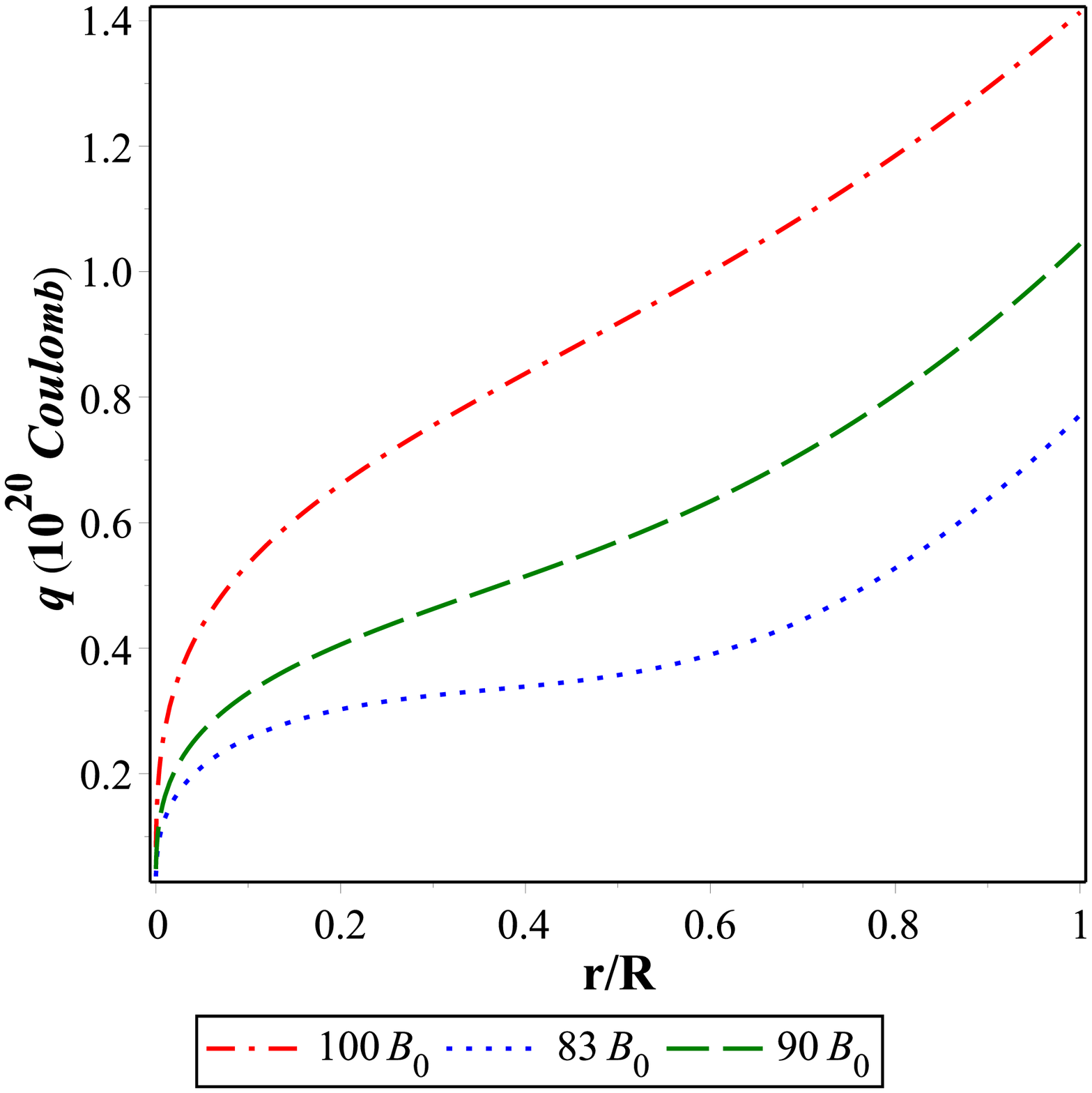}
	\caption{Variation of the density (upper left panel), pressure (upper right panel) and charge (lower panel) with respect to the fractional radial coordinate $r/R$ for the strange star $LMC\ X-4$} \label{Fig1}
\end{figure}

The pictorial representation of density, pressure, and charge are shown in Fig.~\ref{Fig1}. From the variation, it is very clear to understand that density and pressure possess singularity but both of them decrease rapidly with the radius of the star. At the surface of the star, they become a minimum. This feature indicates that the interior i.e., the core of the star is highly compact and our model is valid for outside of the core region. We have presented the properties through three graphs for three different consecutive values of the Bag constant i.e., for $83\ MeV/fm^3$, $90\ MeV/fm^3$ and $100\ MeV/fm^3$ in each frame. We have considered these values from the following literature~\cite{Rahaman2014,Kalam2013}. Though a more wide range of this Bag constant also possible according to the result of CERN-SPS and RHIC. The variation of charge with respect to the fractional radial coordinate represents a usual variation. It increases with the radial coordinate of the star and from the pictorial representation, it is very clear that as the value of Bag constant increases, the charge also increases very rapidly. So without any hesitation, one can say that more the value of Bag constant represents more charge content of the strange star candidate. It is a very vital and interesting result. Also, the total charge content must satisfy the condition $Q^{2}<2MR$ according to the logarithmic principle. A more dense star requires a greater Bag constant represents density-dependent Bag constant. Bordbar et al.~\cite{Bordbar2012} employed Bag constant which is dependent on density to represent magnetized strange quark star.

\section{Boundary Conditions to Determine the Constants}
As we know the pressure at the surface is zero which leads to the following relation
\begin{eqnarray}
& & {Q}^{2}= \bigg[R \big( 16 B\pi {R}^{3}n+16 B\pi {R}^{3}+4 M{n}^{2}-2 R{n}^{2} \nonumber\\
& & \hspace*{1.2cm} +6 Mn-3 Rn+2 M \big)\bigg]\bigg/\bigg[2 {n}^{2}+3 n+1\bigg],\label{eq25}
\end{eqnarray}
where $M$ is the total Mass content of the ultradense compact star.

We can match our interior solution with the exterior Reissner-Nordstr{\"o}m metric as in Eq. (\ref{eq18}). Matching of the component $g_{tt}(r)$ and $e^{\lambda(r)}$ provides us the expressions of the model parameters $K$ and $n$, given by 
\begin{eqnarray}\label{eq24}
& & K= \left( 1-{\frac {2 M}{R}}+{\frac {{Q}^{2}}{{R}^{2}}}
 \right)  \left( {\frac {1}{R}} \right) ^{2n},\nonumber \\
& & n=\big[112B\pi {R}^{3}+6M-9R+( 18688{B}^{2}{\pi }^{2}{R}^{6}- \nonumber \\
& & \hspace{0.5cm} 3264BM\pi {R}^{3} -2016B\pi {R}^{4}+36{M}^{2}\nonumber \\
& & \hspace{0.5cm} -108MR+81{R}^{2} )^{\frac{1}{2}}\big]\big/ 8\big[4B\pi {R}^{3}-3\big].
\end{eqnarray}

Here we have assumed values of the total mass ($M$), bag constant ($B$) and hence have derived the values of the constants, like the total charge ($Q$), $K$ and $n$ in the above expressions. The numerical values of this parameter $n$ is very important because Tolman~\cite{Tolman1939} found out the physically valid solutions for $n=0.5$. We have provided the 
values of this parameter in Table~\ref{Table2}. From the table we can define the characteristic features of this important parameter $n$.

The numerical values of this parameter is very important because Tolman~\cite{Tolman1939} found out the physically valid solutions for $n=0.5$. We have provided the values of this parameter in Table~\ref{Table2}. From this table we can define the character of this important parameter.

\section{Physical Features of the Proposed Model}
Now we will study the physical features of our isotropic, charged fluid configuration in the Einstein-Maxwell space time in the following subsections.

\subsection{Energy Conditions}

For the physical validity of the stellar configuration an isotropic charged fluid sphere composed of Strange Quark Matter (SQM) should satisfy energy conditions, viz., null energy condition (NEC), weak energy condition (WEC), strong energy condition (SEQ) and dominant energy condition (DEC) at all the interior points of the system~\cite{Maurya2017} are given as follows:
\begin{eqnarray}
NEC&:& \rho+p \geq 0,\label{eq31} \\
WEC&:& \rho+p \geq 0, \rho+\frac{E^2}{8\pi} \geq 0,\label{eq32} \\
SEC&:& \rho+p \geq 0, \rho+3p+\frac{E^2}{4\pi} \geq 0,\label{eq33} \\
DEC&:& \rho+\frac{E^2}{8\pi} \geq 0, \rho-p+\frac{E^2}{4\pi} \geq 0. \label{eq34}
\end{eqnarray}

\begin{figure}[!htpb]\centering
\includegraphics[width=6cm]{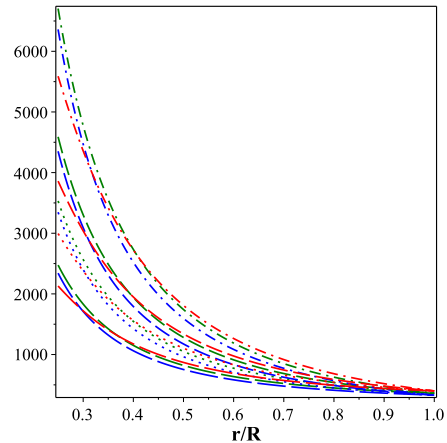}
\caption{Variation of the energy conditions with respect to the fractional radial coordinate $r/R$ for the strange star candidate $LMC\ X-4$. Here, dashdot, dash, dot and longdash linestyle represent $\rho+3p+\frac{E^2}{4\pi}$, $\rho+p$, $\rho+\frac{E^2}{8\pi}$ and $\rho-p+\frac{E^2}{4\pi}$, respectively, whereas blue, green and red colours represent cases with $B=83\ MeV/fm^3$, $B=90\ MeV/fm^3$ and $B=100\ MeV/fm^3$ respectively} \label{Fig2}
\end{figure}

We have represented the energy conditions for the strange star candidate $LMC\ X-4$, with the values of the Bag constant $83\ MeV/fm^3$, $90\ MeV/fm^3$ and $100\ MeV/fm^3$. The graphical variation in Fig.~\ref{Fig2} says that the energy conditions are satisfied and well consistent by our model for the chosen values of  $B$.


\subsection{Generalized TOV Equation}
In order to study compact stars, we must holistically first investigate the stellar structure of these objects. The stellar properties of any compact object are dependent on it’s internal structure, which is described by the Equation of State (EoS). The stability of the compact stellar system can be obtained when the inward gravitational force should be counter balanced by the repulsive and outward forces produced inside the stellar object in such a way that the resultant force on the system would be zero.  Otherwise a small perturbation will cause the system to be unstable. Using the Einstein field equations Oppenheimer and Volkoff \cite{11} derived a differential equation using the work of Tolman \cite{22} that describes the stellar structure of a compact object of static, isotropic material in hydrostatic equilibrium. However, in case of charged compact stars, an extended TOV equation was found first by Bekenstein \cite{33} and thereafter that was followed by other authors \cite{44,55,66,Maurya2017},  as
\begin{equation}
-\frac{M_G(\rho+p)}{r^{2}}e^{\frac{\lambda-\nu}{2}}-\frac{dp}{dr}+\frac{q}{4\pi\,r^4}\,\frac{dq}{dr}=0,\label{eq35}
\end{equation}
where $M_G=M_G(r)$ is the effective gravitational mass inside a sphere of radius $r$. We can derive $M_G(r)$ from the modified
Tolman-Whittaker formula~\cite{Devitt1989} as
\begin{equation}
M_G(r)=\frac{1}{2}r^{2}e^{\frac{\nu-\lambda}{2}}\nu'.\label{eq36}
\end{equation}

Using Eq.~(\ref{eq36}) in Eq.~(\ref{eq35}) we get the following form of TOV equation
\begin{equation}
-\frac{(\rho+p)\nu'}{2}-\frac{dp}{dr}+\frac{q}{4\pi\,r^4}\,\frac{dq}{dr}=0. \label{eq37}
\end{equation}

The above equation represents the equilibrium conditions of an isotropic charged fluid sphere under the combined effect of gravitational, hydrostatic and electrostatic forces:
\begin{equation}
F_g+F_h+F_e=0. \label{eq38}
\end{equation}
where $F_g$ represents gravitational force $\big(\frac{\nu^{\prime}}{2}(\rho+p)\big)$, $F_h$ is the hydrostatic force $\big(\frac{dp}{dr}\big)$ and $F_e$ signifies electrostatic force $\big(\frac{q}{4\pi\,r^4}\frac{dq}{dr}\big)$. 

\begin{figure}[!htpb]\centering
\includegraphics[width=6cm]{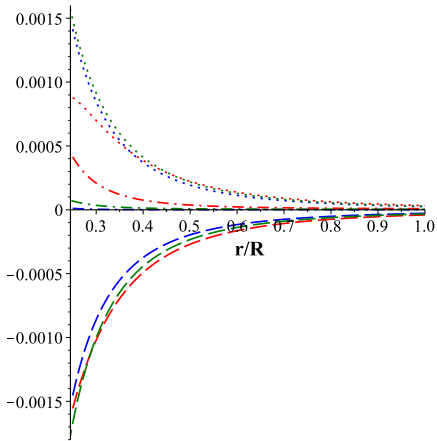}
\caption{Variation of the different forces with respect to the fractional radial coordinate $r/R$ for the strange star candidate $LMC\ X-4$. Here, dot, dashdot and dash linestyle represent $F_h$, $F_e$ and $F_g$, respectively, whereas blue, green and red colours represent cases with $B=83B_0, 90 B_0$ and $100 B_0$, respectively} \label{Fig3}
\end{figure}

The variation of the above forces with respect to the radial coordinate $r/R$ for different values of $B$ are shown in Fig.~\ref{Fig3}. Now, if one looks at Fig. 3  then it will be very much clear that for $B=83B_0$, the positive side of the graph contains $F_h$ and $F_e$ whereas the negative portion of the graph contains $F_g$. Here the force $F_g$ is balanced by the joint action of the two forces $F_h$ and $F_e$ which means that the combine effect of $F_h$, $F_e$ and $F_g$ gives us zero value, and hence physically balanced situation. This suggests that our considered system is in hydrostatic equilibrium. Similarly, we have also checked the validity of our consideration for $B=90B_0$, B=90Bo and $B=100B_0$.
So our model is in stable equilibrium under different forces.


\subsection{Adiabatic Index}
Ratio of the two specific heat, known as adiabatic index $\Gamma$, represents the stiffness of the equation of state for a given density.
It also signifies the stability of fluid sphere which may be relativistic and non relativistic. Using this adiabatic index one can study
dynamical stability of the stellar structure against an infinitesimal radial adiabatic perturbation~\cite{Chandrasekhar1964,Bardeen1966,Wald1984,Knutsen1988,Hererra1997,Horvat2011,Doneva2012,Mak2013,Silva2015}. The following literature survey~\cite{Heintzmann1975,Hillebrandt1976,Chan1993}, tells us that for a stable Newtonian sphere the adiabatic index should exceed $4/3$ inside a dynamical stable fluid distribution. We can write $\Gamma$ for our system as
\begin{equation}
\Gamma=\frac{p+\rho}{p}\frac{dp}{d\rho}. \label{eq39}
\end{equation}


\begin{figure}[!htpb]\centering
\includegraphics[width=6cm]{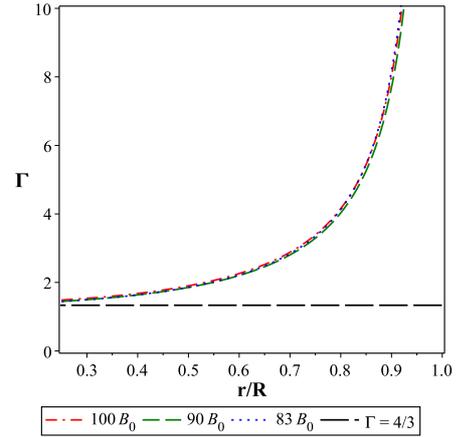}
	\caption{Variation of the adiabatic index with respect to the fractional radial coordinate $r/R$ for the strange star $LMC\ X-4$ } \label{Fig4}
\end{figure}


The graphical variation of adiabatic index in Fig.~\ref{Fig4} signifies the validity of our proposed model for isotropic charged strange star.
The results are true for each chosen values of Bag constant.

\subsection{Mass-Radius Relationship}
Buchdahl~\cite{Buchdahl1959} provided a precise restriction on the upper bound of the mass to radius ratio in uncharged perfect fluid model and it should be $\frac{2M}{R}<\frac{8}{9}$. Its charged generalization i.e. lower bound for a charged compact object was developed by B\"{o}hmer and Harko~\cite{Bohmer2007} which was given by
\begin{equation}
\frac{Q^{4}+18R^{2}Q^{2}}{12R^{4}+R^{2}Q^{2}} \leq \frac{2M}{R}.\label{eq40}
\end{equation}
Andr\'{e}asson~\cite{Andreasson2009} find out the upper bound of the mass to radius ratio for charged fluid sphere and it can be written as
\begin{equation}
\frac{2M}{R} \leq \frac{2\left(2R^2+3Q^2+2R\sqrt{R^2+3Q^2}\right)}{9R^2}.\label{eq41}
\end{equation}
So $\frac{2M}{R}$ must satisfy the following inequality
\begin{eqnarray}
&\qquad\hspace{-0.8cm}\frac{Q^{4}+18R^{2}Q^{2}}{12R^{4}+R^{2}Q^{2}} \leq \frac{2M}{R} \leq \frac{2\left(2R^2+3Q^2+2R\sqrt{R^2+3Q^2}\right)}{9R^2}.\label{eq42}
\end{eqnarray}

The effective gravitational mass for a spherically symmetric isotropic charged stellar system can be given by
\begin{equation}
M_{eff}=\int _{0}^{R}4\pi \left(\rho(r)+\frac{E^2}{8\pi}\right)r^2 dr,\label{eq43}
\end{equation}
where $\left(\rho(r)+\frac{E^2}{8\pi}\right)$ is the effective density for the charged case.

\begin{figure}[!htpb]\centering
\includegraphics[width=8cm]{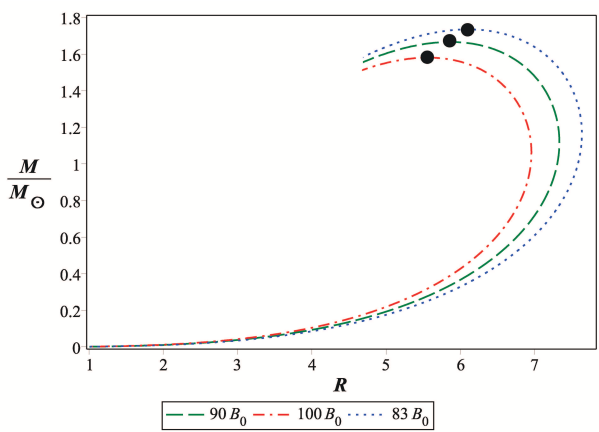}
	\caption{Variation of the $M/M_\odot$ with respect to the radius of the strange star $LMC\ X-4$. Here, the solid dots represent the maxmimum mass points for the strange stars} \label{Fig5}
\end{figure}

In Fig.~\ref{Fig5}, variation of the total mass $M$ (normalized in solar mass), i.e. $\frac{M}{M_\odot}$, with respect to the radius of the strange star is presented very clearly. The solid circles represents the maximum mass point on each curve. It is very clear from the graph that maximum value of mass gradually decreases for increasing value of Bag constants. So as Bag constants increases the strange stars become more compact and smaller is the radius, which will make density more intense than a ordinary compact star. Also if we look at the (Table~\ref{Table2}) then it will be crystal clear to us that as the Bag Constant increases for the strange star candidate $LMC~X-4$, its radius gets more smaller than the previous one. For $83\ MeV/{fm^3}$, $90\ MeV/{fm^3}$ and $100\ MeV/{fm^3}$ the radius are $7.603\pm0.027~Km$, $7.267\pm0.040~Km$ and $6.829\pm0.061~Km$ respectively. The increasing nature of Bag Constant reduces the radius of the ultracompact objects.


\subsection{Redshift}
For a static, spherically symmetric, perfect fluid star the term $\frac{M_{eff}}{R}$ is known as compactification factor and it classifies the stellar objects into different categories as follows ~\cite{Jotania2006}, (1) normal star: $M/R\sim 10^{-5}$, (2) white dwarf: $M/R\sim 10^{-3}$, (3) neutron star: $10^{-1}<M/R<1/4$, (4) ultra dense compact star: $1/4<M/R<1/2$ and (5) black hole: $M/R=1/2$.

The compactification factor can be represented as
\begin{equation}
u=\frac{M_{eff}}{R}.\label{eq44}
\end{equation}

The gravitational redshift function can be presented as
\begin{equation}
Z=e^{-\frac{\nu(r)}{2}}-1.\label{eq45}
\end{equation}


\begin{figure}[!htpb]\centering
\includegraphics[width=6cm]{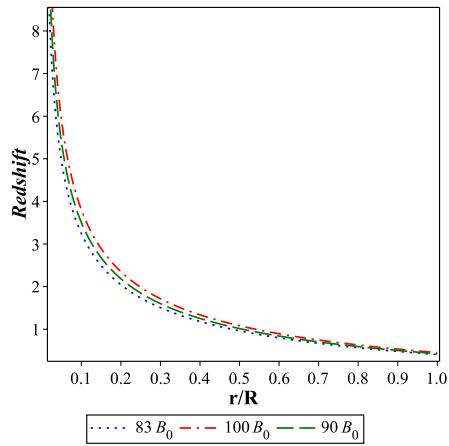}
	\caption{Variation of the gravitational redshift with respect to the fractional radial coordinate $r/R$ for the strange star $LMC\ X-4$} \label{Fig6}
\end{figure}


Barraco and Hamity~\cite{Barraco2002} showed that for an isotropic star redshift must be $\leq 2$ provided the Cosmological constant is absent. After that B\"{o}hmer and Harko~\cite{Bohmer2006} generalizes the above result for anisotropic case and with Cosmological constant and they find out that $Z_{s} \leq 5$. But Ivanov~\cite{Ivanov2002} in one of his pioneering work showed that modification of restriction leads the maximum acceptable value equal to $5.211$. From the Table~\ref{Table2} it is very clear to all of us that its value always remain $\leq 1$. So our model represents a stable model for ultradense strange star.

\begin{table*}[!htp]
\setlength\tabcolsep{15pt} \centering \caption{Physical parameters of the strange star candidates for $B=90\ MeV/{fm^3}$}
      \begin{tabular}{@{}llllllll@{}}
\hline\hline  Strange & Observed & Predicted & $Q$ & Values  & $\frac{2M}{R}$ & Surface    \\
                     Stars   & Mass $(M_\odot)$    & Radius (Km)   & (coulomb)     & of  n   & &   Redshift    \\
\hline         $EXO~1785-248$        & $1.3\pm0.2$       &$7.259\pm0.191$   &$1.144\times 10^{20}$    &0.50      &0.53         & 0.46    \\
               $LMC~X-4$                  & $1.29\pm0.05$   & $7.267\pm0.040$   &$1.078\times 10^{20}$   &0.50       &0.52         & 0.44    \\
               $SMC~X-1$                  & $1.04\pm0.09$   & $7.321\pm0.029$   &$1.333\times 10^{20}$   &0.42      &0.42         & 0.31    \\
               $SAX~J~1808.4-3658$  & $0.9\pm0.3$       & $7.235\pm0.280$   &$1.676\times 10^{20}$   &0.38      &0.37       & 0.26    \\
               $4U~1538-52$              & $0.87\pm0.07$    & $7.207\pm0.072$   &$1.725\times 10^{20}$   &0.37      &0.36       & 0.25    \\
              $HER~X-1$                  & $0.85\pm0.15$    &$7.185\pm0.169$    &$1.752\times 10^{20}$   &0.36       &0.35       &0.24    \\
\hline\hline \label{Table1}
\end{tabular}
\end{table*}

\begin{table*}[!htp]
\setlength\tabcolsep{12pt} \centering \centering \caption{Physical parameters of the strange star candidate $LMC\ X-4$}
\begin{tabular}{ccccccccc}
\hline\hline Case &  Values of  & Predicted  & $Q$  & Values  & $\frac{2M}{R}$   &  Surface    \\
                  &   B $(MeV/{fm^3})$ & Radius (Km) & (coulomb) & of $n$  &      &Redshift     \\
\hline I & 83                             &$7.603\pm0.027$   &$0.698\times 10^{20}$    &0.48      &0.50         & 0.41\\
       II &  90                             &$7.267\pm0.040$   &$1.078\times 10^{20}$    &0.50      &0.52         & 0.44  \\
       III &  100                           &$6.829\pm0.061$   &$1.414\times 10^{20}$    &0.52      &0.56         & 0.51 \\

\hline\hline \label{Table2}
\end{tabular}
\end{table*}

\section{Concluding Remarks}
Strange stars are always an interesting object of study for the astrophysicists. Because each an every point of view regarding strange star
opens up a new feature of it as the layers of an onion. In one of our previous work, we have already studied the strange star with Tolman $V$ metric potential~\cite{Shee2018}. In that paper, we have studied the situation without charge and clarify the stability of it. Here we generalized the situation with charge. More extensively here we studied charged, spherically symmetric, the isotropic strange star with Tolman $V$ metric potential. Consideration of more realistic EOS of Quantum Chromo Dynamics makes the study more prominent, pinpointed and complicated~\cite{Alford2013}.

$\textbf{1. Density, Pressure, and Charge:}$
Solutions of Einstein Field Equations provide us the expression for density $({\rho})$ and pressure $(p)$ as in Eq.~(\ref{eq19}) and Eq.~(\ref{eq20}). From the matching condition of the component ${g_{tt}}$ for our system and the exterior Reissner-Nordstr{\"o}m metric provide us the expression of the total charge content $(Q)$ of our system in Eq.~(\ref{eq26}). The central density and the central pressure both suffer the singularity problem and their graphical nature Fig.~\ref{Fig1} provide us the information that with the increase of the radius of the star these physical features decreases rapidly and significantly. The variation of the charge content of the system with the radius of the star is finite, increases monotonically and maintaining its nature for differently chosen values of Bag Constant. From Fig.~\ref{Fig1} it is also clear that as Bag Constant increases the charge content also increases significantly. It is a very important result for further study. In Table~\ref{Table1} we have listed the numerical value of the total charge content of the different strange star candidates for Bag Constant $90\ MeV/{fm^3}$. It is  $\sim10^{20}$ coulomb for each of these stars. Also, we have studied the total charge content of a particular strange star candidate $LMC\ X-4$ with the different values of Bag Constant in Table~\ref{Table2}. Though the order is the same but total charge content increases with the increasing values of the Bag Constant respectively.

$\textbf{2. Value of n:}$
We have found out the expression of different constants by matching our space-time with the exterior Reissner-Nordstr{\"o}m metric. With the help of this matching, we have found out the expression of $n$ which is very important regarding the study of this paper. Because Tolman originally used $n=\frac{1}{2}$ for physically acceptable solutions. From our model we have find out the values of $n$ for $B=90\ MeV/{fm^3}$ as $0.50, 0.50, 0.42, 0.38, 0.37$ and $0.36$ for the Strange star candidates  $EXO~1785-248$, $LMC~X-4$ , $SMC~X-1$, $SAX~J~1808.4-3658$, $4U~1538-52$ and $HER~X-1$ respectively (Table~\ref{Table1}). Also for $LMC~X-4$nwe have find out the value of $n$ with Bag Constant $83\ MeV/{fm^3}$, $90\ MeV/{fm^3}$ and $100\ MeV/{fm^3}$ in Table~\ref{Table2}. The corresponding values are $0.50$, $0.52$ and $0.56$ respectively. So it is very clear that as the Bag Constant increases the value of $n$ also increasing in nature. This is a case of further study for researchers.Very recently Aziz et al.~\cite{Aziz2019} obtained from different observational data for several compact objects that the possibility of the wide range of Bag constant should be $(41.58 - 319.31) MeV/fm^3$.

$\textbf{3. Energy conditions:}$
Our charged isotropic fluid distribution satisfies all the energy conditions viz. NEC, WEC, SEC and DEC as in Eq.~(\ref{eq29}) - Eq.~(\ref{eq32}). We have studied these equations for the strange star candidate $LMC\ X-4$ for three consecutive value of Bag constant i.e. $83\ MeV/{fm^3}$, $90\ MeV/{fm^3}$ and $100\ MeV/{fm^3}$ respectively. Fig.~\ref{Fig2} tells us that all the energy conditions having a maximum value at the center and reduces gradually towards the surface of the star.

$\textbf{4. TOV equation:}$
The generalized TOV equation for our static, isotropic, charged, spherically symmetric fluid sphere is given in Eq.~(\ref{eq35}). The balancing nature of our model under gravitational, hydrostatic and electrostatic forces is given in Fig.~\ref{Fig3}. We have represented the graphs for three consecutive values of the Bag Constant i.e. $83\ MeV/{fm^3}$, $90\ MeV/{fm^3}$ and $100\ MeV/{fm^3}$ respectively.

$\textbf{5. Adiabatic Index:}$
Another stability checking criterion is the adiabatic index. For a stable, spherically symmetric fluid sphere the adiabatic index should be greater than $\frac{4}{3}$. From Fig.~\ref{Fig4} it is clear that our proposed model satisfies the required criteria for Bag Constant $83\ MeV/{fm^3}$, $90\ MeV/{fm^3}$ and $100\ MeV/{fm^3}$ respectively. So in respect to the adiabatic index, we have proposed a stable model.

$\textbf{6. Surface Redshift:}$
In Table~\ref{Table1} we have found out surface redshift of different strange star candidates for Bag Constant $90\ MeV/{fm^3}$. A critical analysis tells us that these values are much less than $1$. So it provides a stable structure. Also for the strange star candidate $LMC~X-4$ (Table~\ref{Table2}) we have find out the surface redshift for Bag Constant  $83\ MeV/{fm^3}$, $90\ MeV/{fm^3}$ and $100\ MeV/{fm^3}$ which are $0.41$, $0.44$ and $0.51$ respectively. So it is very clear that as Bag Constant increases the value of surface redshift also increases.

The overall study of this paper suggests that though the density and pressure suffer the problem of singularity but the other stability criteria satisfied by our proposed model. So our charged stellar model with SQM distribution and Tolman $V$ metric potential serves a new path for the researchers to study the ultradense compact stars.

\section*{Acknowledgement}
SR is thankful to the Inter University Centre for Astronomy and Astrophysics (IUCAA) for providing Visiting Associateship under which a part of this work has been carried out. SR is also thankful to the Authority of The Institute of Mathematical Sciences, Chennai, India for providing all types of working facility and hospitality under Associateship scheme. DD is also grateful to the authority of IUCAA for the warm hospitality and library facility during the visit. SKM and MKJ acknowledges support from the authority of University of Nizwa, Nizwa, Sultanate of Oman.

\end{document}